\documentclass[prl,twocolumn,showpacs,superscriptaddress]{revtex4-1}
\usepackage{graphicx,dcolumn,bm,color,amsmath,amssymb}
\newcommand{\lcvo}{$\rm LiCuVO_4$}

\begin{document}
\title{Suppression of 3D-ordering by defects in the $S=1/2$ frustrated chain magnet LiCuVO$_4$}

\author{L.\,A.\,Prozorova}
\author{S.\,S.\,Sosin}
\email{sosin@kapitza.ras.ru}
\author{L.\,E.\,Svistov}
\email{svistov@kapitza.ras.ru}
\affiliation{P.L. Kapitza Institute for Physical Problems RAS, 119334 Moscow, Russia}
\author{N.\,B\"{u}ttgen}
\affiliation{Center for Electronic Correlations and Magnetism (EKM), Experimentalphysik V, Universit\"{a}t Augsburg, D--86135 Augsburg, Germany}
\author{J.\,B.\,Kemper}
\author{A.\,P.\,Reyes}
\author{S.\,Riggs}
\affiliation{National High Magnetic Field Laboratory, Tallahassee, FL 32310, USA}
\author{O.\,A.\,Petrenko}
\affiliation{Department of Physics, University of Warwick, Coventry CV4 7AL, UK}
\author{A.\,Prokofiev}
\affiliation{Institut f\"{u}r Festk\"{o}rperphysik Technische Universit\"{a}t Wien, A--1040 Wien, Austria}

\date{\today}

\begin{abstract}
We report on a heat capacity study of high quality single crystal samples of \lcvo\ -- a frustrated spin $S=1/2$ chain system -- in magnetic field amounting to 3/4 of the saturation field. At low fields up to about 7~T, a linear temperature dependence of the specific heat, $C_p\propto T$, resulting from 1D magnetic correlations in the spin chains is followed upon cooling by a sharp lambda anomaly of the transition into a 3D helical phase with $C_p\propto T^3$ behavior at low temperature. The transition from a spin liquid into a spin-modulated (SM) phase at higher fields occurs via a hump-like anomaly which, as the temperature decreases further turns into a $C_p\propto T^2$ law distinctive for a quasi-2D system. We suggest an explanation for how nonmagnetic defects in the Cu$^{2+}$ chains can suppress 3D long-range ordering in the SM phase and leave it undisturbed in a helical phase.
\end{abstract}

\pacs{75.50.Ee, 76.60.-k, 75.10.Jm, 75.10.Pq}
\maketitle

Quantum-spin chains with frustrated exchange interactions were among the most interesting issues for both experimental and theoretical research in condensed matter physics of the past decade \cite{Hikihara_08,Nishimoto_12,Starykh_14}. The enhanced effect of quantum fluctuations imposed upon a fine balance of exchange interactions leads to a variety of novel ground states and phase transformations in these systems \cite{Kecke,Sudan,Sato_09,Zhitomirsky_10}. \lcvo\ is an example of a quasi-1D magnet whose unconventional magnetic phases result from a competition of ferromagnetic and antiferromagnetic exchange interactions between nearest-neighbor ($J_1$) and next-nearest neighbor ($J_2$) in-chain spins. As a result of this particular combination of exchange interactions a helical incommensurate spin structure is stabilized in this system below $T_N\approx 2.3$~K \cite{Gibson_04}. A strong reduction in the ordered spin component of Cu$^{2+}$ ions in this state $\langle\mu\rangle/\mu_B\simeq 0.3$ \cite{Gibson_04,Enderle_05} provides evidence that the system partially retains properties of 1D chains.

Moderate applied magnetic fields of 7-8~T induce a transformation of the spin helix into a collinear spin-modulated (SM) structure in which all the spins are parallel to the field, with their ordered components oscillating along the chain with an incommensurate period.This transition may be related  to the field evolution of short range chiral (transverse) and spin-density wave (longitudinal) correlations for the 1D $J_1$-$J_2$ model \cite{Heidrich_09}.

In the field range just below the saturation field ($\mu_0H_{sat}\simeq 45$~T) \cite{Svistov_11} the theory predicts the presence of a long-range nematic ordering \cite{Chubukov_91,Zhitomirsky_10}. The magnetic properties of \lcvo\ in this field range were studied by pulse magnetization and nuclear magnetic resonance (NMR) techniques \cite{Svistov_11,Buettgen_14}. These experiments specified the supposed field range where the spin nematic phase can exist. The experiments discussed in our report were performed at fields well below this range. This work is solely focused on the magnetic properties of \lcvo\ in the helical and SM phase at lowest temperatures, and the spin-liquid phase for elevated temperatures.

Single crystals of \lcvo\ grown by the methods described in \cite{Prokofiev_04} belong to the same series as the ones studied in our previous high field NMR experiments \cite{Buettgen_14}. The main part of the specific heat measurements were carried out in the Tallahassee National High Magnetic Field laboratory with the use of a home-made relaxation type calorimeter operating at temperatures 1.8--30~K in magnetic field up to 35~T. Supplementary measurements were performed on a Quantum Design Physical Property Measurement System (PPMS) with a $^3$He insert and a 9~T cryomagnet. The samples of 0.67 and 5.8~mg by mass were studied in the first and the second experiment respectively. The magnetic field was applied along the $c$~axis of the crystal.

\begin{figure}
\centerline{\includegraphics[width=0.76\columnwidth]{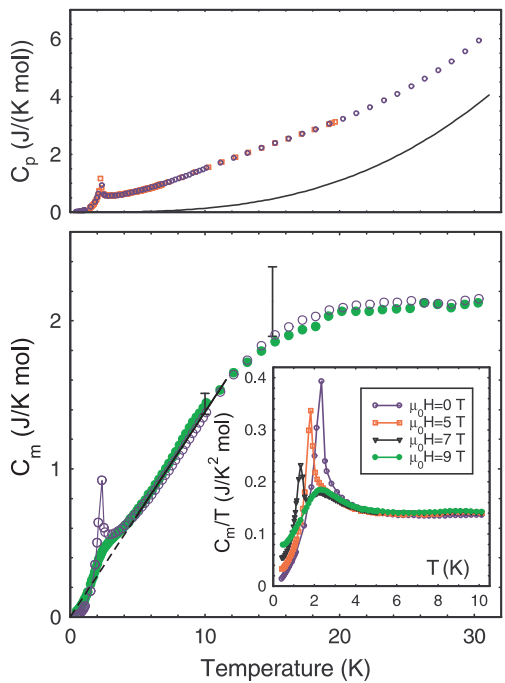}}
\caption{(Color online) Upper panel: temperature dependence of the specific heat at zero magnetic field measured using a PPMS (circles) and using a home made high-field setup (squares); the solid line is an approximation of the phonon contribution to the heat capacity.
Lower panel: magnetic specific heat $C_m(T)$ (phonon contribution subtracted) of the sample measured in a PPMS at $\mu_0H=0$ ({\Large $\circ$}) and 9~T ({\Large $\bullet$}), the error bars are explained in the text, solid line is a linear fit to both curves in the temperature range from 5 to 10~K; the inset shows the specific heat divided by temperature as a function of $T$ obtained at several values of magnetic field (see legend).}
\label{CvsT}
\end{figure}

Zero-field results obtained in both experiments and presented in Fig.~\ref{CvsT} are in perfect agreement with each other.
The phonon contribution to the heat capacity (expressed in units J/(K~mol)) has been approximated by the function $C_p=1.4\cdot 10^{-4} \; T^3$ (solid line in the upper panel). It was chosen so that the magnetic entropy (with the phonon part subtracted) attains the value of $0.7R\ln 2$ at a temperature 30~K in agreement with numerical results for the 1D $J_1$-$J_2$ model \cite{Sirker}. This rough estimation gives a correction to the total heat capacity of about 3\%, 8\% and 20\% at $T=5$, 10 and 15~K, respectively, shown by the error bars in the lower panel of Fig.~\ref{CvsT}. Within these bounds our data conform with previous measurements \cite{Yamaguchi,Smirnov}. This panel shows a comparison between the magnetic part of the specific heat $C_m(T)$ obtained at zero field and at a field of 9~T in the PPMS measurements. No significant difference between the two curves is observed in the high-temperature range. All the data in the lower panel of Fig.~\ref{CvsT} exhibit temperature dependences $C_m\propto T$ in the range $5-10$~K providing evidence for the presence of 1D short-range correlations in the spin chains.
The linear in temperature section of a zero-field curve is followed by a sharp lambda anomaly at the transition into a 3D helical phase at $T_N=(2.27\pm 0.03)$~K in agreement with a number of previous studies \cite{Gibson_04,Enderle_05}.
The inset to the lower panel of Fig.~\ref{CvsT} presenting $C_m/T (T)$ curves at various fields traces the evolution of this transition shifting to low temperature as the field increases. For the field $\mu_0H=9$~T which is greater than the field of the transition into a SM phase a lambda anomaly is replaced by a hump-like feature followed by a $C_m\propto T^2$ dependence towards lower temperatures. This curve slightly deviates from the square law at lowest temperatures.

\begin{figure}
\centerline{\includegraphics[width=0.8\columnwidth]{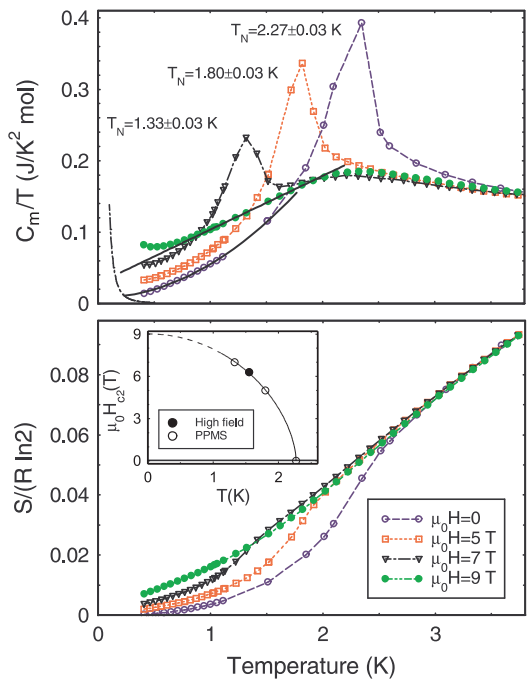}}
\caption{(Color online) Upper panel: Expansion of the low temperature part of the inset to Fig.~1, solid lines are quadratic and linear extrapolations to the zero field and 9~T curves respectively, dashed-dotted line is a calculated contribution from
$^{63}$Cu and $^{65}$Cu nuclei at the field 10~T.
Lower panel: integrated $C_m/T (T)$ curves at various field values normalized to $R\ln{2}$; inset shows the boundary of the helical state obtained from both the PPMS and the high field measurements, the line is a guide-to-the-eye.}
\label{entropy}
\end{figure}

This low-temperature part of the data requires more careful investigation and is expanded in the upper panel of Fig.~\ref{entropy}. The phase transition at zero field is followed by a distinct $C_m\propto T^3$ behavior of the specific heat with a small linear contribution revealing itself at the lowest temperature (corresponding fit is presented in the Fig.~\ref{entropy}). Shifting of the N\'{e}el temperature under magnetic field ($T_N$ values are annotated in the upper panel and shown by open circles in the inset to lower panel) is accompanied by the increase of this low-$T$ contribution.
In the absence of the transition peak at $\mu_0H=9$~T the high-$T$ linear part of the curve is converted into a quasi-2D $C_m\propto T^2$ law extending for half a decade by temperature (see linear fit to $C_m/T$ curve for 9~T). The low-$T$ rise of this curve might indicate the proximity to a 3D transition. Extrapolation of the phase boundary $\mu_0H_{c2} (T_N)$ (see the inset to the lower panel of Fig.~\ref{entropy}) to zero temperature gives the value $\mu_0H_{c2} (0)\simeq 9$~T.
Alternatively, this rise could be attributed to the specific heat of the Cu nuclei (the corresponding curve estimated for the external field 10~T and zero hyperfine field is shown in Fig.~\ref{entropy}). In any case, the total entropy evolved in the process of low-$T$ disordering does not depend on the external field in this field range: the corresponding functions obtained by integration of $C_m/T(T)$ curves converge at temperatures above the ordering transition at zero field.

\begin{figure}
\centerline{\includegraphics[width=0.8\columnwidth]{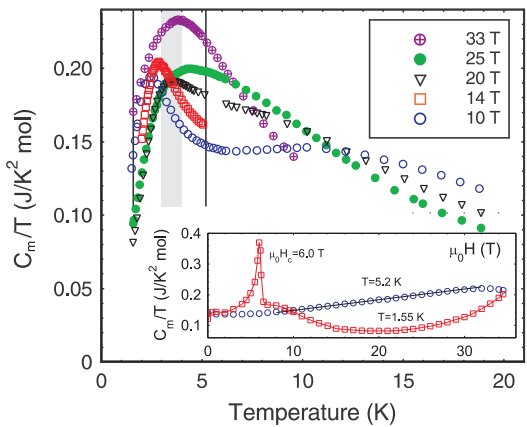}}
\caption{(Color online) Magnetic part of the specific heat divided by temperature as a function of $T$ measured at various magnetic fields using a high field setup; shadowed area corresponds to the formation of static 2D magnetic correlations as observed in an NMR experiment \protect\cite{Buettgen_12}, vertical lines mark temperatures of scans in magnetic field shown in the inset. Solid line in the inset is a linear fit to data at $T=5.2$~K in the intermediate field range.}
\label{C_over_T}
\end{figure}

A further increase in the applied magnetic field leads to a considerable change in the specific heat behavior (see Fig.~\ref{C_over_T}). A broad hump in $C_m/T$ vs $T$ curve preceding the formation of static 2D correlations shifts to somewhat higher temperatures, while the quasi-1D linear part (constant for $C_m/T$) breaks down at fields above 20~T. It is replaced by another broad feature merging with the first one at a maximum experimental field of 33~T. Two scans in an applied magnetic field performed at constant temperature below (1.55~K) and above (5.2~K) the ordering transition $T_N=2.27$~K at $(H=0)$ are presented in the inset to Fig.~\ref{C_over_T}. The low-$T$ curve has a sharp peak associated with the transition from a helical phase at $\mu_0H_{c2}=6.0$~T followed by a gradual decrease in magnitude towards higher fields. The specific heat reaches its minimum around 20~T and then starts to increase as the field enters the range of critical fluctuations near saturation. In contrast, the high-$T$ curve demonstrates a crossover from a nearly constant to a linear field dependence $C/T(H)$ (the fit is shown by solid line) extending to about 25~T. The final downturn of the specific heat developing at fields above 30~T indicates the approach of the system to the saturated state.

We shall now discuss the specific heat data obtained for high-quality single crystal samples of \lcvo\ at field and temperatures of spin-liquid and ordered helical and SM phases. The specific heat in the spin-liquid phase for temperatures $5\lesssim T\lesssim 10$~K is a nearly linear function of $T$ at low fields ($H\lesssim H_{sat}/4$) and becomes nonlinear at higher fields. The observed evolution is in qualitative agreement with numerical results obtained for a Heisenberg $S=1/2$ chain model \cite{Klumper}. The temperature dependence of the specific heat in the ordered helical state can be roughly approximated as a sum of linear and $C_m\propto T^3$ terms while for the SM state the cubic term is replaced by $C_m\propto T^2$ contribution.
Different exponents in power laws point to 3D and quasi 2D properties of these phases respectively confirming that the antiferromagnetic ordering between neighboring $ab$-planes in the SM structure is destroyed. This effect was previously reported in neutron scattering and NMR measurements \cite{Buettgen_10,Enderle_12}. The replacement of a sharp lambda anomaly at the transition to the helical state by a broad feature at the crossover to the SM state gives another hint of this change.

\begin{figure}
\centerline{\includegraphics[width=0.61\columnwidth]{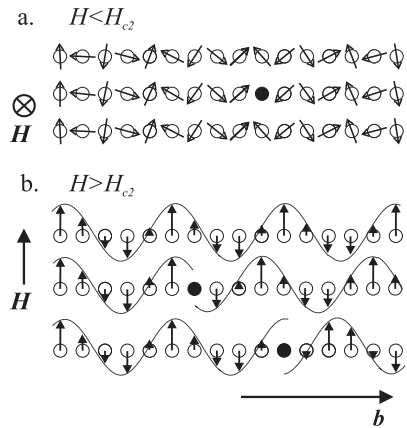}}
\caption{Schematic diagram of the influence of defects on the spiral state (a) and the spin-modulated state at $\mu_0H\simeq 20$~T (b), defects are shown by black dots. The spiral state is nearly insensitive to the chain break while the phase of a spin-modulated chain is randomly shifted at the point of the defect.}
\label{scheme}
\end{figure}
In the following we propose a qualitative model to explain this different dimensional behavior by considering the effect of a nonmagnetic impurity. A composition study of \lcvo\ reveals a Li deficiency of a few percent even in crystals grown using the low temperature technique which provides crystals of the highest perfection. It was argued that the Li deficiency results in holes with spins $S=1/2$ localized on oxygen which in turn form a Zhang-Rice singlet with the neighboring copper spin \cite{Prokofiev_04,Zhang}. Such a singlet should be equivalent to a non-magnetic impurity replacing a Cu spin and will provoke an unusual magnetic state in its vicinity with the two nearest ferromagnetic exchange bonds removed, but with a conserved antiferromagnetic bond between the two parts of the interrupted chain. The presence of such defects in \lcvo\ single crystals of the same series as in this work was suggested in order to explain the bulk  magnetization process, as well as the vanadium NMR spectra in the vicinity of the saturation field \cite{Buettgen_14}. A schematic diagram of helical and SM spin chains broken by a non-magnetic defect is shown in Fig.~\ref{scheme}. In contrast to a spin chain with only a nearest-neighbor exchange interaction, the frustrated $J_1$-$J_2$ chain is not fully broken by a non-magnetic impurity. The magnetic moments on both sides of the defect are coupled by indirect antiferromagnetic exchange interaction leading to a local distortion and phase shift of an incommensurate exchange structure.

Obviously, the presence of this type of defect does not significantly disturb the helical magnetic structure in \lcvo\ since the incommensurate wave vector of the helix ${\rm k_{ic}}=0.468\cdot 2\pi/b$ (the unit cell of \lcvo\ contains two magnetic ions of Cu) corresponds to the angle between next-nearest neighbor spins close to $\pi$ (see Fig.~\ref{scheme}a) and does not depend on the applied field. A residual phase shift of a helical structure near the defect may be eliminated due to an intra-chain exchange interaction so that the initial 3D long range ordering is not suppressed. In contrast, the influence of defects on the SM ordering is expected to be more pronounced for the following reasons. The wave vector of a SM structure in a 1D $J_1$-$J_2$ model is field dependent and expressed as follows: ${\rm k_{ic}}=(1-M(H)/M_{sat})\cdot \pi/b$  \cite{Hikihara_08,Starykh_14,Kecke,Sudan,Sato_09}. This relation has been traced experimentally in the field range below 15~T \cite{Masuda_11,Enderle_12} and therefore, the value ${\rm k_{ic}}$ is significantly less than $\pi /b$ at all fields above $H_{\rm c2}$. The elongation of the spin modulation period along the chain in field leads to a random shift of its phase at each defect (see Fig.~\ref{scheme}b, the scheme of the moment distribution over the chain is presented for $\mu_0H\simeq 20$~T).
The boundary conditions for spins at both ends of a finite chain should depend on its length leading to an uncertainty in the phase shift of an incommensurate structure due to a random distribution of the vacancies. As in case of a helical structure, the effect of these inter-chain exchange interactions is to restore these broken correlations and establish a 3D ordering.
Taking into account the hierarchy of exchange parameters (the interaction between $c$~planes is the smallest one) correlations between spins in neighboring $ab$~planes are hindered and the system remains in an effectively 2D ordered state.

In conclusion, a specific heat experiment was performed on high-quality single crystals of \lcvo. The results show that while a helical spin state observed in the low-field range $H<H_{\rm c2}$ clearly demonstrates 3D magnetic ordering, the field induced magnetic phase at $H>H_{\rm c2}$ has properties peculiar to a quasi-2D system. We propose a simple model to describe how nonmagnetic defects in the Cu$^{2+}$ chains can destroy 3D long-range ordering in a SM phase, but leave it undisturbed in the helical phase. In addition, the proposed model allows one to simultaneously explain the results of both NMR \cite{Buettgen_12,Nawa_13,Buettgen_14} and neutron scattering measurements \cite{Enderle_12} resolving the existing discrepancy in their interpretations. According to NMR data, the magnetic phases at both sides of $H_{\rm c2}$ are dipolar so that the average values of magnetic moments of Cu ions are non-zero on a time scale of NMR measurements ($\sim 1$~sec). This proves the presence of static magnetic correlations and precludes an identification of the spin-nematic state, assumed in \cite{Enderle_12} for higher magnetic fields $H>H_{\rm c2}$. Nevertheless, the short-range character of these correlation especially in the direction perpendicular to $ab$ planes detected in neutron diffraction patterns can be understood using the above model describing the suppression of a 3D long range order by non-magnetic defects. Further specific heat experiments extending to the vicinity of the saturation field (about 45~T) is especially interesting in prospect of possible observation of the transition from a SM into a theoretically predicted quadrupole nematic state.

This work is supported by the Grants 13-02-00637 of the Russian Fund for Basic Research, Program of Russian Scientific Schools and by the German Research Society (DFG) via TRR80. The authors thank M.~Zhitomirsky for stimulating discussions.

\end{document}